\documentclass[aps,prl,reprint,superscriptaddress]{revtex4-1}
\usepackage{hyperref}
\usepackage{amsmath}
\usepackage{amssymb}
\usepackage[pdftex]{graphicx}
\usepackage{microtype}
\usepackage{verbatim}
\usepackage{todonotes}
\presetkeys{todonotes}{inline,backgroundcolor=yellow}{}
\usepackage{subcaption}
\makeatletter
\DeclareRobustCommand{\element}[1]{\@element#1\@nil}
\def\@element#1#2\@nil{%
  #1%
  \if\relax#2\relax\else\MakeLowercase{#2}\fi}
\makeatother

\newcommand{\mrm}{\mathrm}

\newcommand{\bp}[1]{\left( #1 \right)} 
\newcommand{\ba}[1]{\left< #1 \right>} 

\newcommand{\ket}[1]{{\left| {#1} \right\rangle}}
\newcommand{\bra}[1]{{\left\langle {#1} \right|}}

\newcommand{\braket}[2]{{\left\langle {#1}|{#2} \right\rangle}}

\newcommand{\Esca}{\mathcal{E}}

\begin{document}
\title{Improved estimate of the collisional frequency shift in Al$^+$ optical clocks}
\author{Jack Davis}
\affiliation{Department of Physics, University of Toronto, Toronto, Canada M5S 1A7}
\author{Pierre Dub\'{e}}
\affiliation{National Research Council Canada, Ottawa, Canada K1A 0R6}
\author{Amar C. Vutha}
\affiliation{Department of Physics, University of Toronto, Toronto, Canada M5S 1A7}

\begin{abstract}
Collisions between background gas particles and the trapped ion in an atomic clock can subtly shift the frequency of the clock transition. The uncertainty in the correction for this effect makes a significant contribution to the total systematic uncertainty budget of trapped-ion clocks. Using a non-perturbative analytic framework that was developed for this problem, we estimate the frequency shift in Al$^+$ ion clocks due to collisions with helium and hydrogen. Our calculations significantly improve the uncertainties in the collisional shift coefficients, and show that the collisional frequency shifts for Al$^+$ are zero to within uncertainty. 
\end{abstract}
\maketitle

The present generation of optical atomic clocks, using neutral atoms in optical lattices or atomic ions in ion traps, are the most stable timekeepers that have ever been constructed \cite{Ludlow2015}. Of all the trapped-ion optical clocks in operation around the world \cite{Barwood2014,Huntemann2016,Dube2014,Dube2018}, the Al$^+$ clock at NIST \cite{Chou2010,Chou2010PRL,Chen2017,Brewer2019} is currently the world's most accurate. Recent improvements in its accuracy, due to a reduction in the uncertainty from the blackbody radiation shift and the second-order Doppler shift \cite{Chen2017,Brewer2019}, have resulted in accuracy better than 10$^{-18}$. Among the effects that contribute to its residual systematic uncertainty, the collisional frequency shift (CFS) is an important one. The CFS arises from collisions between the clock ion and residual background gas particles in the vacuum chamber. Before this work, the best estimate of the CFS for the Al$^+$ clock had an associated uncertainty of $0.5 \times 10^{-18}$ \cite{Chou2010,Chou2010PRL}, obtained by conservatively assigning the maximum differential phase shift of $\frac{\pi}{2}$ between the ground ($^1S_0$) and excited ($^3P_0$) states of the clock transition per collision. (The fractional frequency uncertainty due to collisional effects in the Al$^+$ clock was recently re-evaluated as $0.24 \times 10^{-18}$, out of which the contribution of scattering phase shifts -- the focus of the present work -- is $0.23 \times 10^{-18}$ \cite{Brewer2019,Hankin2019}.) Improved methods to evaluate the CFS are essential, so that the CFS does not stand in the way of continued improvements to clock performance.

Evaluation of the CFS with improved accuracy requires knowledge of the scattering phase shifts (or equivalently, scattering amplitudes) in the potential energy curves associated with each of the clock states during collisions with background gas particles. The dominant background gas species in the ultra-high-vacuum environment of a trapped-ion clock are typically hydrogen molecules and helium atoms. The required ground and excited potential energy curves for exotic systems such as AlHe$^+$ and AlH$_2^+$ (which we shall refer to as ``molecules'' in the following) must in general be obtained from \emph{ab initio} calculations. The scattering phase shifts must also be combined with appropriate weights, since the collision cross sections (and therefore the collision rates) also depend on the scattering potentials, which are generally quite different for the two clock states.

In Ref. \cite{vutha2017collisional}, a quantum-channel description of the collision between a clock ion and background gas particles was used to develop a master equation, which allows the CFS to be evaluated in a straightforward manner. These calculations were limited to collisions between clock ions and helium atoms for simplicity; however, the predominant background gas in trapped-ion clock systems is molecular hydrogen. In this work, we significantly extend the methods developed in Ref.\ \cite{vutha2017collisional} and apply it to a problem of immediate relevance: we develop a master equation that includes both unitary and non-unitary effects of collisions, and use it to calculate the CFS for Al$^+$ clock ions colliding with hydrogen molecules and helium atoms. Our results significantly reduce the systematic uncertainty associated with the CFS for the Al$^+$ clock.

\textit{Analytic framework.} We briefly review the essential steps involved in calculating the CFS. The effect of a collision can be described by considering the unitary dynamics of the clock ion and background gas particle during the collision process, followed by a trace over the background gas degrees of freedom. We model the Al$^+$ clock ion as a two-level system with ${}^1S_0$ and ${}^3P_0$ states. Since the hyperfine interaction in the two clock states is extremely weak, we assume that the nuclear spin degree of freedom of ${}^{27}$Al$^+$ is decoupled from the problem. Throughout this paper, we use units where $\hbar=1$ for convenience. 

For the elastic collisions that we consider in this work, the effect of a collision on the clock ion's density matrix is described by a set of Lindblad jump operators $L_\ell$ associated with each partial wave collision channel $\ell$, and a mean field Hamiltonian $\mathcal{H}_M$. The matrix elements of these operators in the clock ion state space are (see Supplementary Material, Section A)
\begin{equation}\label{eq:matrix_elements}
\begin{split}
[L_\ell]_{\alpha \beta} & = \delta_{\alpha \beta} \, \sqrt{\frac{4 \pi}{k^2} (2\ell+1)} \, |\sin \phi_{\ell,\alpha}| \, e^{i \phi_{\ell,\alpha}} \\
[\mathcal{H}_M]_{\alpha \beta} & = -\delta_{\alpha \beta} \, \left( \frac{\pi \, n_\mrm{bg} v}{k^2} \right) \sum_\ell (2\ell + 1) \, \sin 2\phi_{\ell,\alpha}.
\end{split}
\end{equation}
Here $\alpha,\beta \in \{g,e\}$ are indices labeling the clock states, $\phi_{\ell,\alpha}$ are the $\ell$-th partial wave scattering phase shifts for the clock state $\ket{\alpha}$. The collision energy is $k^2/2\mu$, $\mu$ is the reduced mass of the colliding particles, $v=k/\mu$ is their relative speed, and $n_\mrm{bg}$ is the number density of the background gas.

The Lindblad jump operators and mean field Hamiltonian enter the master equation for the density matrix of the clock states: $\frac{d\rho}{dt} = -i[H_0 + \mathcal{H}_M,\rho] + \sum_\ell L_\ell \rho L_\ell^\dagger - \frac{1}{2}\sum_\ell \left\{ L_\ell^\dagger L_\ell, \rho \right\}$, where the terms involving $L_\ell$ describe the dissipative dynamics due to the collision, and $H_0$ is the Hamiltonian for unitary dynamics due to, e.g., the trapping potential, probe laser, etc. During their time evolution under this equation, the off-diagonal density matrix elements incur extra phase rates compared to their collision-free evolution, which can be identified with the CFS (see Supplementary Material, Section B). The resulting expression for the CFS correction is 
\begin{equation}\label{eq:frequency_shifts}
\begin{split}
\delta \omega_\mrm{CFS} & = n_\mrm{bg} v  \frac{4\pi}{k^2}  \sum_\ell (2\ell + 1) \left( A_\ell + B_\ell \right); \\
A_\ell & = \frac{1}{4} \left( \sin2\phi_{\ell,e} - \sin2\phi_{\ell,g} \right), \\
B_\ell & = |\sin \phi_{\ell,e} \, \sin\phi_{\ell,g}| \, \sin(\phi_{\ell,e} - \phi_{\ell,g}), 
\end{split}
\end{equation}
where we define $\delta \omega_\mrm{CFS} = \omega_0 - \omega_m$, with $\omega_0$ the unperturbed resonance frequency and $\omega_m$ the resonance frequency measured in the presence of collisions. The $A_\ell$ terms are the shift of the clock frequency due to the mean field correction $\mathcal{H}_M$, while the $B_\ell$ terms originate from the dissipative part of the master equation described by the jump operators $L_\ell$.  

\textit{Numerical results.}  The phase shifts required to evaluate the CFS from Eq.\ (\ref{eq:frequency_shifts}) were calculated in the following way. Potential energy curves (PECs) for the AlHe$^+$ and AlH$_2^+$ molecules were calculated using the PSI4 package \cite{Parrish2017}, with \texttt{cc-pVTZ} basis sets \cite{Woon1994} for all the atoms. To obtain potential energy curves that are adiabatically connected to the ground ($^1S_0$) and excited ($^3P_0$) clock states, the equation of motion coupled cluster (EOM-CCSD) method \cite{stanton1993equation,Wang2014} was used, as implemented in PSI4. For AlH$_2^+$, the separation between the H atoms was fixed at 1.45 $a_0$ \cite{Alexander2007} for all the energy calculations. Energy eigenvalues were evaluated at separations between the Al$^+$ ion and the background gas particle ranging from $r = 2 \, a_0$ to $r = 50 \, a_0$, and the results were interpolated using cubic splines to yield continuous PECs. 

\begin{figure}
\includegraphics[width=\columnwidth]{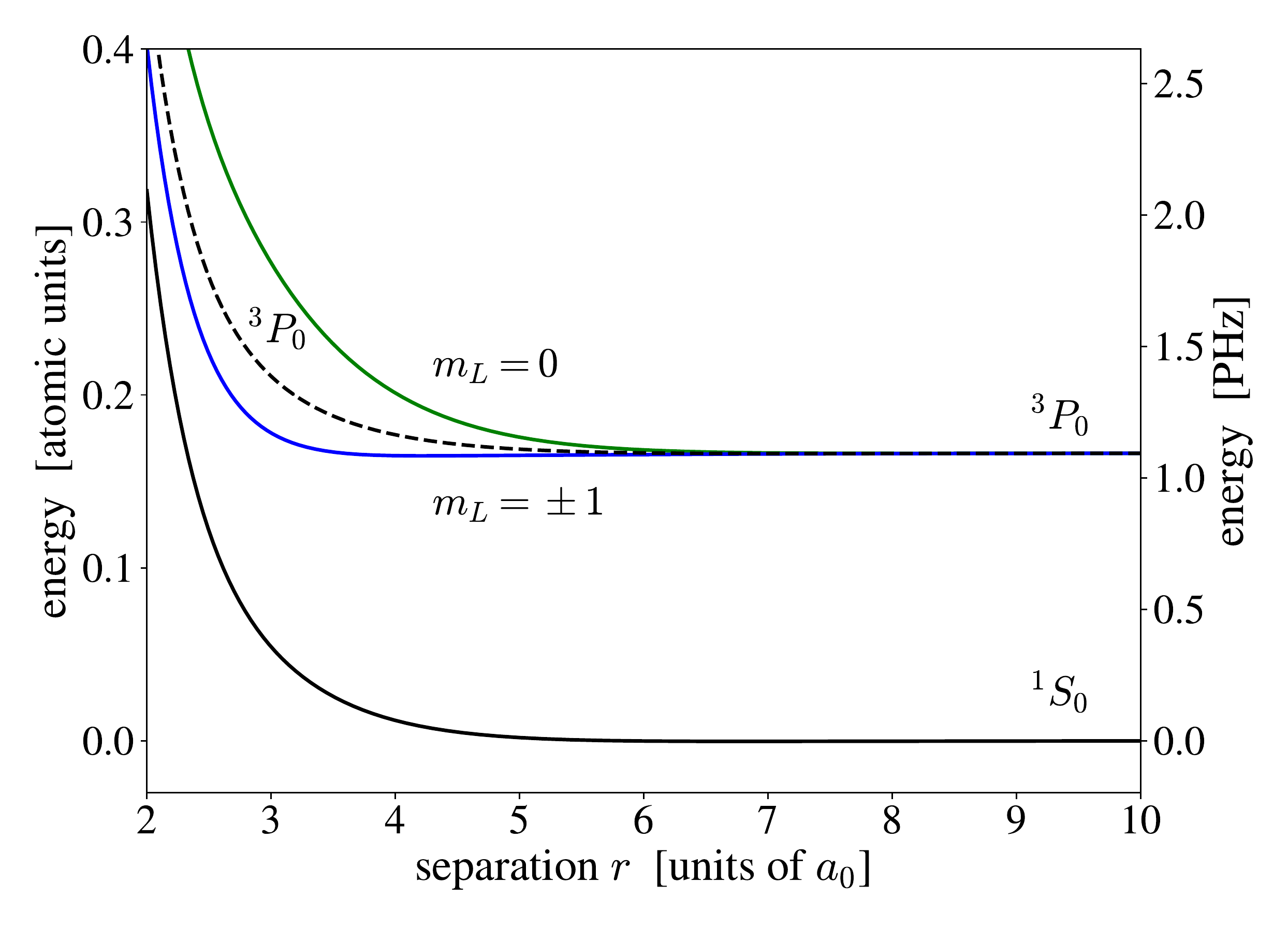}
\caption{Numerically calculated potential energy curves for the AlHe$^+$ molecule. The excited state potential energy curves for $m_L = 0,\pm1$ are averaged with equal weights to obtain an estimate of the potential energy curve for the $^3P_0$ clock state (dashed curve).}
\label{fig:he-pecs}
\end{figure}

Despite some recent progress in \emph{ab initio} methods \cite{epifanovsky2015spin}, it remains challenging to compute excited-state PECs that fully account for spin-orbit interactions. The PSI4 package does not implement spin-orbit coupling, and therefore the energy levels we obtained using the EOM-CCSD method correspond to different azimuthal quantum numbers ($m_L = 0,\pm 1$) of the $3s 3p$ wavefunction with respect to the collision axis (also the quantization axis), rather than the spin-orbit-coupled ${}^3P_{0,1,2}$ levels. In order to compute the molecular PEC connected to the ${}^3P_0$ excited clock state for subsequent scattering calculations, we used the Clebsch-Gordan decomposition of the ${}^3P_0$ state,
\begin{equation}
\begin{split}
\ket{{}^3P_0} & = \sum_{m_L,m_S} C_{m_L \, m_S} \ket{L,m_L} \ket{S,m_S} \nonumber \\
& = \frac{1}{\sqrt{3}} \big( \ket{1,1}\ket{1,-1} - \ket{1,0}\ket{1,0} + \ket{1,-1}\ket{1,1} \big).
\end{split}
\end{equation}
Since the $\ket{m_L = 0,\pm1}$ states each contribute with equal probability to $\ket{{}^3P_0}$, we used the average of the $m_L = 0,\pm1$ PECs as a reasonable estimate of the correct PEC for the ${}^3P_0$ state. 
This procedure leads to the potential curves shown in Fig.\ \ref{fig:he-pecs} for the AlHe$^+$ molecule.

\begin{figure}
\includegraphics[width=\columnwidth]{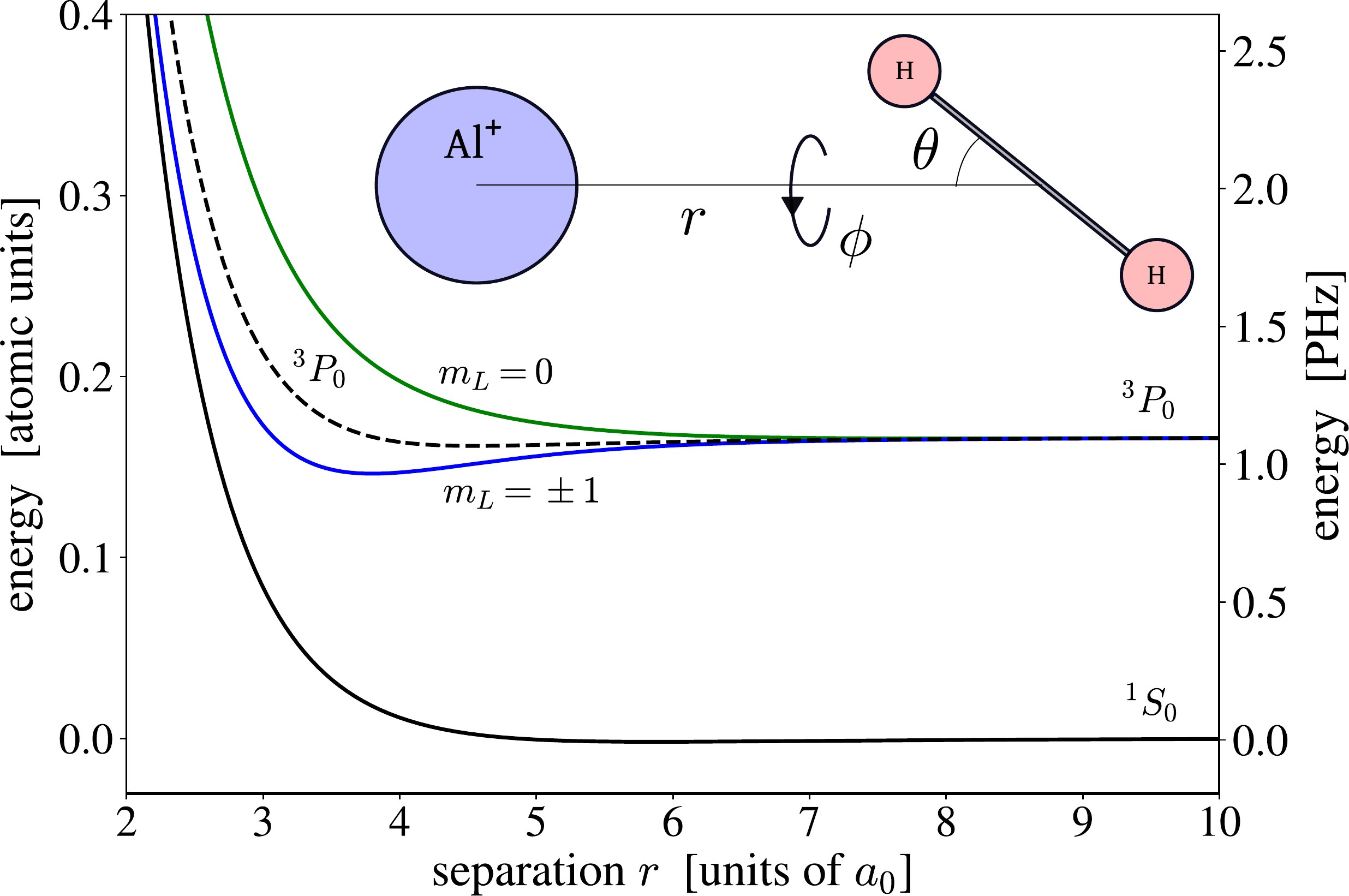}
\caption{Energy eigenvalues of the AlH$_2^+$ molecule were calculated for fixed values of $r,\theta,\phi$, (geometry as shown in the inset) and spherically averaged to obtain the potential energy curves. The resulting curves for $m_L = 0,\pm1$ are averaged to obtain the ${}^3P_0$ potential energy curve (dashed line), as with helium.}\label{fig:h2-pecs}
\end{figure}

For Al$^+$ -- H$_2$ collisions, the rotational degree of freedom of the hydrogen molecule needs to be considered. We reduced the resulting potential energy surfaces to potential energy curves by averaging over the orientation of the H$_2$ molecule, using the following procedure. Since the H$_2$ molecules are drawn from a thermal bath, each of the $2J+1$ $m_J$-sublevels $\ket{J,m_J}$ of a rotational state of the molecule have identical populations. As a result, the probability distribution for the orientation of the molecular axis is a uniform distribution over a sphere. We calculated the energy eigenvalues for a set of polar angles $\theta\in\ \{ \frac{\pi}{8}, \frac{\pi}{4}, \frac{3\pi}{8}, \frac{511\pi}{1024} \}$ (where $\theta$ is defined as shown in Fig.\ \ref{fig:h2-pecs}), for each value of the Al$^+$ -- H$_2$ separation $r$. By smoothly connecting the resulting energies as a function of $r$, we obtained a set of PECs for each value of $\theta$. Since the interaction of Al$^+$ and H$_2$ is symmetric in the azimuthal angle $\phi$, the PECs for different values of $\theta$ were averaged together with $\sin \theta$ weight factors to obtain the spherically averaged potential curves shown in Fig.\ \ref{fig:h2-pecs}. 

\begin{figure}
        \includegraphics[width=\columnwidth]{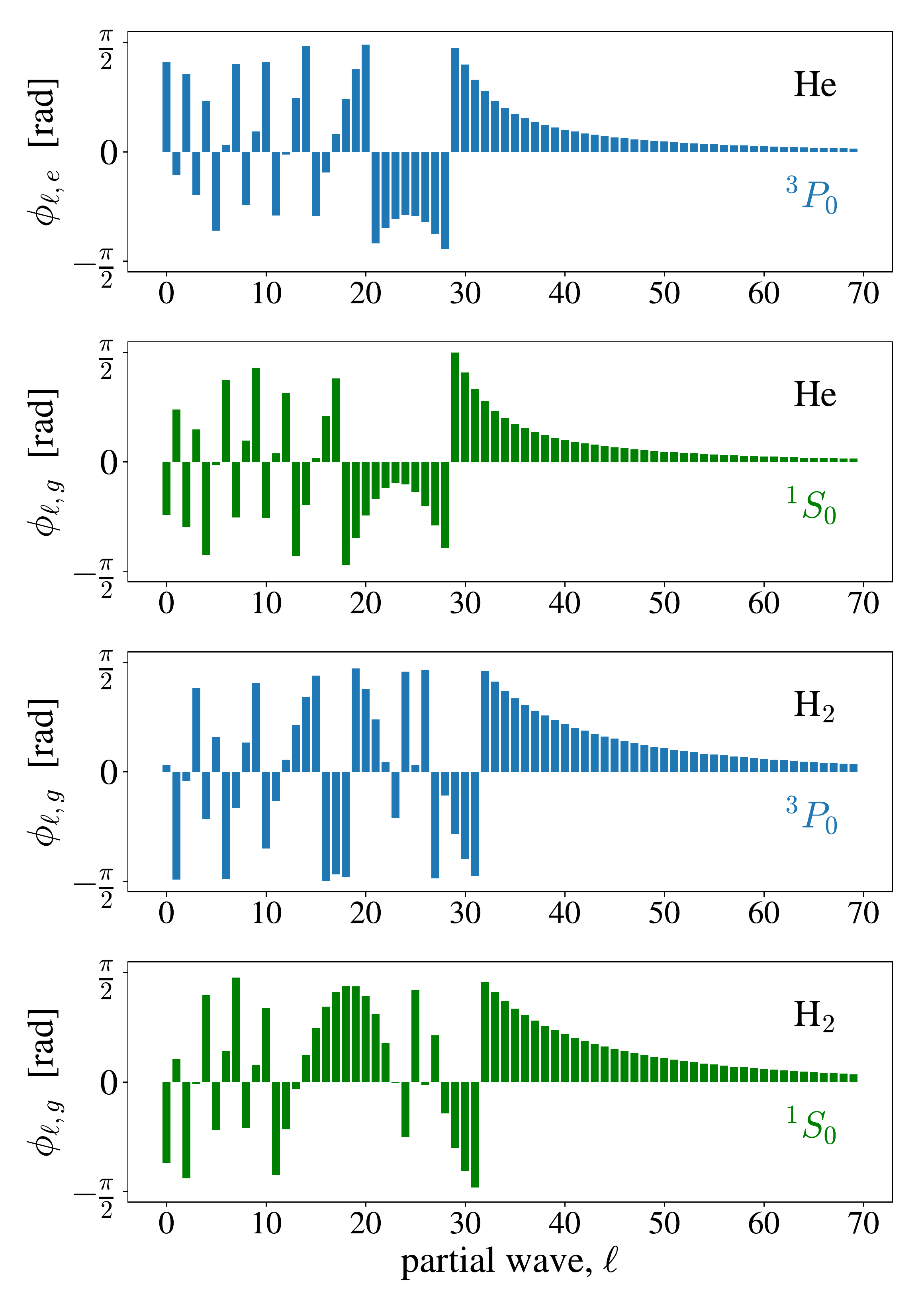}
    \caption{Partial wave phase shifts for scattering of Al$^+$ with He and H$_2$, at a collision energy of 295 K. The upper two plots show phase shifts for the clock states under collisions with He atoms, and the lower two plots are phase shifts for collisions with H$_2$ molecules.}
    \label{fig:phase-shifts}
\end{figure}

\begin{figure}[hp!]
\centering
\includegraphics[width=\columnwidth]{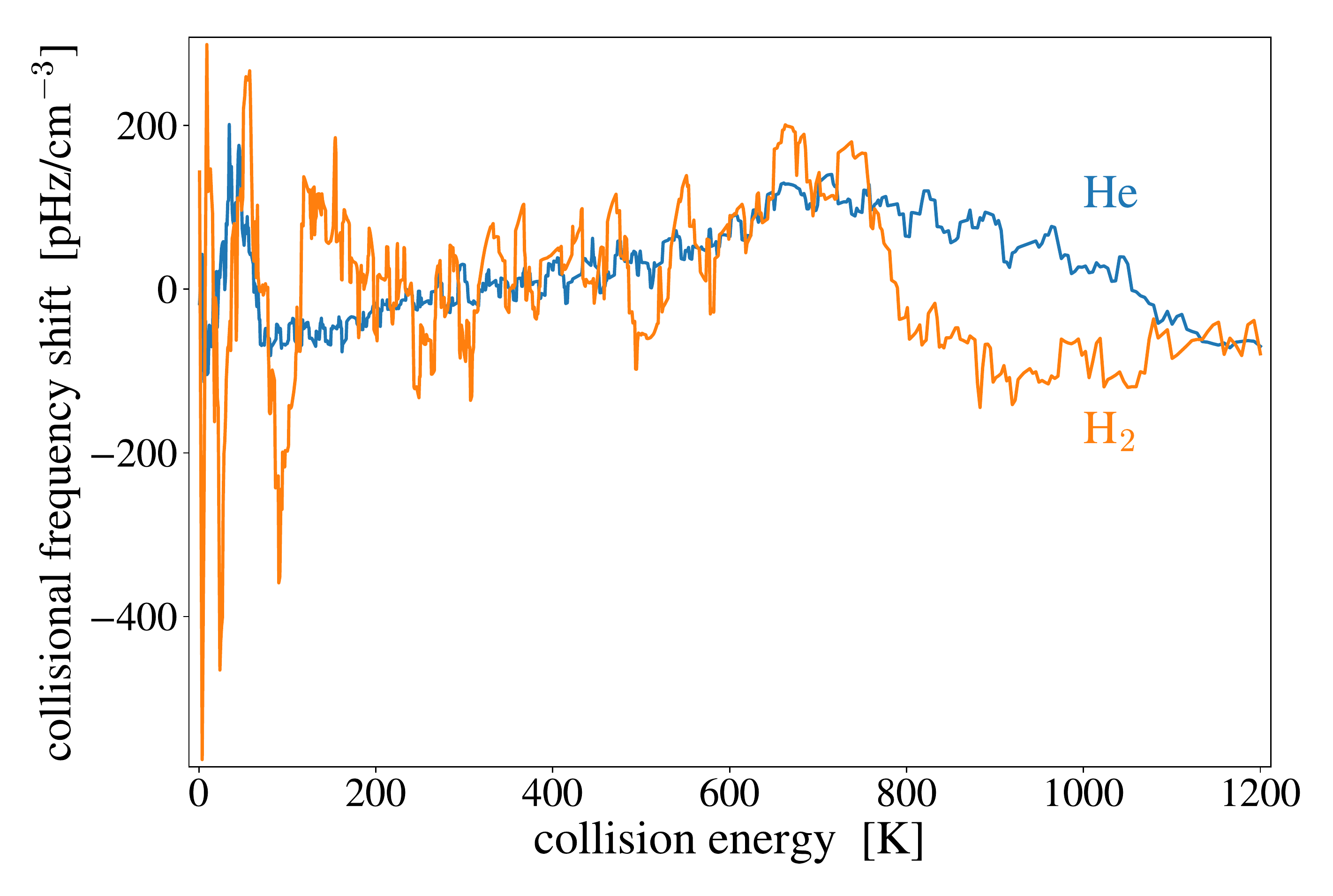}
\caption{Collisional frequency shifts for the Al$^+$ clock transition as a function of collision energy, for He and H$_2$ scattering.}
\label{fig:cfs-v-energy}
\end{figure}

Scattering wavefunctions were obtained by numerical integration of the Schr\"{o}dinger equation for each PEC. The values of $\phi_{\ell,\alpha}$ (where $\alpha$ labels the PEC) were extracted using the formula \cite{Joachain1975} 
\begin{equation}\label{eq:phase-shift-arctan}
\phi_{\ell,\alpha} = \tan^{-1}\left[ \frac{k j_\ell'(kr_0) - \beta_\ell j_\ell(kr_0)}{kn_\ell'(kr_0) - \beta_\ell n_\ell(kr_0)} \right],
\end{equation}
where $j_\ell$ ($n_\ell$) are spherical Bessel (Neumann) functions and $\beta_\ell = \left[ R'_\ell(r)/R_\ell(r) \right]_{r=r_0} $ is the log derivative of the radial eigenfunction, $R_{\ell,s}(r)$, evaluated at $r_0$. The phase shifts were computed with $r_0 = 50 \, a_0$, much larger than the range of the potentials, so that the phase shifts could be extracted accurately. A typical distribution of the resulting partial wave phase shifts for a collision energy of 295 K is shown in Fig.\ \ref{fig:phase-shifts}. The scattering phase shifts were computed for PECs adiabatically connected to the ground (${}^1S_0$) and excited (${}^3P_0$) clock states, for collision energies ranging from 1 to 1200 K, and for 100 partial waves per collision energy. The resulting collisional frequency shifts, as a function of collision energy, are shown in Fig.\ \ref{fig:cfs-v-energy}. Thermally averaged CFS values were obtained by performing Boltzmann averages over the collision energy, with bath temperatures of 295 K (representing a room-temperature clock apparatus) and 10 K (in consideration of future cryogenic optical clocks).

The largest systematic uncertainty in the CFS calculation stems from the neglect of spin-orbit interaction in the \emph{ab initio} potential curves for the excited clock state. We conservatively assign the systematic uncertainty in our CFS calculations to be the maximum difference between the CFS computed with the ($m_L$-averaged) $^3P_0$ PEC, and the CFS computed using the individual $3s3p \  m_L = 0,\pm1$ PECs. At 295 K, the thermally averaged CFS is
\begin{equation*}
\begin{split}
\ba{\delta\omega}_{\text{He}} & = 2\pi \times \bp{ 14 \pm 32 } \, \text{pHz} \times \left[\frac{n_\mrm{bg}}{\mrm{cm}^{-3}} \right], \\
\ba{\delta\omega}_{\text{H}_2} & = 2\pi \times \bp{12 \pm 161 } \, \text{pHz} \times \left[\frac{n_\mrm{bg}}{\mrm{cm}^{-3}} \right].
\end{split}
\end{equation*}
The shifts are zero to within uncertainty, and the uncertainties in the collisional shift coefficients (CFS per unit background gas density) are markedly smaller than the estimates used for Al$^+$ clocks to date \cite{Chou2010,Chou2010PRL,Hankin2019}. With $n_\mrm{bg} = 2.7 \times 10^5$ cm$^{-3}$ (corresponding to a reference pressure level of 1 nPa at 295 K), the fractional frequency uncertainty in the CFS estimated for helium (hydrogen) collisions is $8 \times 10^{-21}$ ($4 \times 10^{-20}$). 

At 10 K, the thermally averaged CFS is
\begin{equation*}
\begin{split}
\ba{\delta\omega}_{\text{He}} & = 2\pi  \times \bp{ -7.0 \pm 18 } \, \text{pHz} \times \left[\frac{n_\mrm{bg}}{\mrm{cm}^{-3}} \right], \\
\ba{\delta\omega}_{\text{H}_2} & = 2\pi  \times \bp{-34 \pm 66 } \, \text{pHz} \times \left[\frac{n_\mrm{bg}}{\mrm{cm}^{-3}} \right].
\end{split}
\end{equation*}
The collisional shift coefficients at 10 K are not very different from the room temperature coefficients. Improvements to the CFS in cryogenic trapped-ion clocks, compared to room-temperature clocks, are therefore likely to result from improved vacuum levels in a cryogenic environment, rather than any strong temperature dependence of the CFS coefficients.

We have restricted our attention here to elastic collisions of the trapped ion with background gas particles. Inelastic collisions that transfer the ion out of the subspace spanned by the two clock states can limit its interaction time with the probe laser, and degrade the signal-to-noise ratio of the clock, but do not result in frequency shifts (cf. \cite{Dorscher2018} for a related analysis for photon scattering). On the other hand, collisions that are inelastic in the internal states of background gas particles (e.g., H$_2$) could affect the scattering phase shifts that enter the CFS calculations. Vibrational excitations of H$_2$ are frozen out at all the relevant temperatures, but rotational transitions are possible in principle. We estimated the probability for population transfer between the $J=0$ and $J=2$ states in H$_2$ due to the electric field gradient from the ion, by numerically solving for the time evolution of the rotational states along the classical trajectory of the collision (see Supplementary Material, Section C): the rotational excitation probability is $< 1\%$  even for head-on collisions at the collision energies relevant to this work, which justifies our focus on elastic collisions here.



In summary, Al$^+$ ion collisions with He atoms and H$_2$ molecules have been considered, and the resulting collisional frequency shifts calculated with improved accuracy. Our work establishes a systematic method for estimating collisional frequency shifts in optical clocks, which can be applied to other trapped-ion optical clocks that are currently in operation around the world.

~\\ ~ \\
\emph{Acknowledgments.} We acknowledge helpful discussions with David Leibrandt and Shira Jackson. We are grateful to Tom Kirchner for stimulating discussions throughout the course of this investigation. This work is supported by the Branco Weiss Fellowship, the Sloan Fellowship and Canada Research Chairs.

\bibliography{collisions}

\newpage
\onecolumngrid

\section{Supplementary Material}

\subsection*{A. Lindblad jump operators and mean field Hamiltonian for collisions}
We will construct the density matrix equation of motion using a set of Kraus operators \cite{Nielsen2010}, obtained by projecting the scattered wavefunction onto a basis of orbital angular momentum eigenstates (partial waves) for the relative motion degree of freedom of the colliding particles. These Kraus operators will lead us to a set of Lindblad jump operators and a mean field Hamiltonian acting on the internal states of the clock ion. The construction is along similar lines as Ref.\ \cite{vutha2017collisional}, and follows the approach laid out by Preskill \cite{Preskill}. 

The S-matrix for the collision acts on both the relative motion and clock ion internal degrees of freedom. We write it as $S = \mathbb{I} + i T$, where $T$ is the on-shell T-matrix. The matrix element of $S$ between angular momentum eigenstates results in an operator that acts only on the clock states, whose elements are
\begin{equation}
[\bra{\ell} S \ket{\ell'}]_{\alpha \beta}  = \delta_{\alpha \beta} \, \delta_{\ell \ell'} \, e^{i2 \phi_{\ell,\alpha}}.
\end{equation}
Here $\alpha,\beta$ are indices denoting the internal states of the clock ion and take values in $\{g,e\}$. We will also need the overlap of an incident plane wave $\ket{k}$ with momentum $k$ with an outgoing spherical wave (partial wave) $\ket{\ell}$, given by $\braket{\ell}{k} = \sqrt{\frac{\pi}{k^2} (2\ell + 1) \, n_\mrm{bg} v \, \delta t} = \sqrt{\frac{\pi n_\mrm{bg}}{\mu k} (2\ell + 1) \, \delta t}$. Here $n_\mrm{bg}$ is the background gas density, $\mu$ is the reduced mass of the colliding particles, $v$ is the relative velocity of the collision, and $\delta t$ is a coarse-graining timescale (long compared to the duration of a collision, but short compared to the internal dynamics of the clock ion). The amplitude of the incident plane wave is chosen here to be $\sqrt{n_\mrm{bg} v \, \delta t}$, corresponding to a choice of normalization to one particle per unit area.

We can now evaluate the required Kraus operators. With each partial wave $\ell$ we associate a Kraus operator $K_\ell = \bra{\ell} T \ket{k}$ operating on the internal degrees of freedom of the clock ion. This represents the effect of the collision on the clock ion, conditioned on scattering into an outgoing spherical wave with angular momentum $\ell$. The matrix elements of $K_\ell$, in the space spanned by the clock ion internal states, are
\begin{equation}
\begin{split}
[K_\ell]_{\alpha \beta} & =  \sum_{\ell'} [\bra{\ell} T \ket{\ell'}]_{\alpha \beta} \, \braket{\ell'}{k} \\
	   & = \delta_{\alpha \beta} \, e^{i\phi_{\ell,\alpha}} \, \sin \phi_{\ell,\alpha} \, \sqrt{\frac{4\pi}{k^2} (2\ell + 1) \, n_\mrm{bg} v \, \delta t}.
\end{split}
\end{equation}
We also define the ``no-scattering'' Kraus operator $K_\emptyset = \bra{k} S \ket{k}$, whose matrix elements in the internal state space are
\begin{equation}
\begin{split}
[K_\emptyset]_{\alpha \beta} & =  \sum_{\ell \, \ell'} \braket{k}{\ell} \, [\bra{\ell} S \ket{\ell'}]_{\alpha \beta} \,  \braket{\ell'}{k} \\
	   & = \delta_{\alpha \beta} \left\{ 1 + n_\mrm{bg} v \, \delta t \frac{\pi}{k^2} \sum_\ell (2\ell + 1) \, \left[ \left(-2 \sin^2 \phi_{\ell,\alpha} \right) + i \sin 2\phi_{\ell,\alpha} \right] \right\}.
\end{split}
\end{equation}
(We note that $K_\emptyset$ was derived incorrectly in Ref.\ \cite{vutha2017collisional}, leading to a neglect of the imaginary term.)

The Kraus operators can be conveniently rewritten in terms of the scattering rates and scattering amplitudes using standard partial wave expansions \cite{Joachain1975}.
\begin{equation}
\begin{split}
[K_\ell]_{\alpha \beta} & = \delta_{\alpha \beta} \, e^{i \phi_{\ell,\alpha}} \sqrt{\gamma_{\ell,\alpha} \, \delta t} \\
[K_\emptyset]_{\alpha \beta} & = \delta_{\alpha \beta} \left[ 1 - \frac{\gamma_\alpha}{2} \delta t  + i \left(\frac{2\pi n_\mrm{bg}}{\mu} \right) \mathfrak{Re} f_\alpha(0) \, \delta t \right] = \delta_{\alpha \beta} \left[ 1 - \frac{\gamma_\alpha}{2} \delta t - i \Lambda_\alpha \delta t \right], 
\end{split}
\end{equation}
where $\gamma_{\ell,\alpha} = n_\mrm{bg} v \frac{4\pi}{k^2} (2\ell + 1) \sin^2 \phi_{\ell,\alpha}$ is the $\ell$-th partial wave scattering rate, $\gamma_\alpha = \sum_\ell \gamma_{\ell,\alpha}$ is the total scattering rate, and $f_\alpha(0)$ the forward scattering amplitude corresponding to the internal state $\ket{\alpha}$. In the last line, we have also defined the quantities $\Lambda_\alpha = - n_\mrm{bg} v \frac{\pi}{k^2} \sum_\ell (2\ell + 1) \, \sin 2\phi_{\ell,\alpha}$ for convenience. It is easy to verify that the set of Kraus operators satisfies the completeness relation, $K_\emptyset^\dagger K_\emptyset + \sum_\ell K_\ell^\dagger K_\ell = \mathbb{I}$, up to $\mathcal{O}(\delta t^2)$. This ensures that the dynamics of the ion's reduced density matrix, after tracing over the motional degree of freedom, is trace-preserving. 

The density matrix for the internal states of the clock ion evolves due to scattering over the time interval $\delta t$ as
\begin{equation}
\begin{split}
\rho(t + \delta t) & = K_\emptyset \, \rho(t) K_\emptyset^\dagger + \sum_\ell K_\ell \, \rho(t) K_\ell^\dagger \\
& = \rho(t) - i [\mathcal{H}_M, \rho(t)] \, \delta t + \left( \sum_\ell L_\ell \rho(t) L_\ell^\dagger - \frac{1}{2}\sum_\ell \left\{ L_\ell^\dagger L_\ell, \rho(t) \right\} \right) \delta t + \mathcal{O}(\delta t^2).
\end{split}
\end{equation}
This allows us to read off the jump operators $L_\ell = K_\ell/\sqrt{\delta t}$, and the mean field Hamiltonian $[\mathcal{H}_M]_{\alpha \beta} = \delta_{\alpha \beta} \Lambda_\alpha$, which leads to the matrix elements given in Equation (\ref{eq:matrix_elements}). 

Making the Markovian assumption for the bath of background gas particles at this point, and taking the limit $\delta t \to 0$, allow us to write the time evolution of the density matrix as a first-order differential equation in Lindblad form:
\begin{equation}\label{eq:master_equation}
\frac{d\rho}{dt} = -i [H_0 + \mathcal{H}_M, \rho] + \sum_\ell L_\ell \rho L_\ell^\dagger - \frac{1}{2}\sum_\ell \left\{ L_\ell^\dagger L_\ell, \rho \right\}. 
\end{equation}
Here we have also included the unitary time evolution of the clock states under the Hamiltonian $H_0$, which contains the effect of everything other than the collisions (e.g., the trapping potential, probe laser, etc.).

\subsection*{B. Collisional frequency shift in terms of scattering phase shifts}
Making the rotating-wave approximation for the clock laser-ion interaction, the Hamiltonian $H_0$ in matrix form is
\begin{equation}
H_0 = \frac{1}{2}\begin{pmatrix} \Delta & \Omega \\ \Omega & -\Delta \end{pmatrix},
\end{equation}
where $\Delta = \omega-\omega_0$ is the detuning, $\omega$ is the laser frequency, $\omega_0$ the clock ion's resonance frequency and $\Omega$ is the Rabi frequency for the laser-ion interaction. The equation of motion for the off-diagonal density matrix element $\rho_{ge}$, from Eq.(\ref{eq:master_equation}), is then
\begin{equation}
\begin{split}
\frac{d\rho_{ge}}{dt} = & -i \left[\Delta + (\Lambda_g - \Lambda_e)  - \sum_\ell \sqrt{\gamma_{\ell,g} \gamma_{\ell,e}} \sin(\phi_{\ell,g} - \phi_{\ell,e})   \right] \rho_{ge} \\
& + \Omega (\rho_{ee} - \rho_{gg}) - \left[ \frac{(\gamma_g + \gamma_e)}{2}  -\sum_\ell \sqrt{\gamma_{\ell,g} \gamma_{\ell,e}} \cos(\phi_{\ell,g} - \phi_{\ell,e}) \right] \rho_{ge}
\end{split}
\end{equation}
The real terms on the right-hand side affect the amplitude of the coherence $\rho_{ge}$, whereas the imaginary terms lead to a phase shift. The resonance frequency $\omega_m$ is the value of the laser frequency $\omega$ for which there is no phase shift acquired by $\rho_{ge}$. So it is easy to read off $\omega_m$ by setting the imaginary part of the above equation to zero, which leads to the CFS correction
\begin{equation}
\delta \omega_\mrm{CFS} = \omega_0 - \omega_m = -(\Lambda_e - \Lambda_g) + \sum_\ell \sqrt{\gamma_{\ell,g} \gamma_{\ell,e}} \sin(\phi_{\ell,e} - \phi_{\ell,g}).
\end{equation}
Rewriting $\Lambda_\alpha$ and $\gamma_{\ell,\alpha}$ in terms of the scattering phase shifts $\phi_{\ell,\alpha}$ results in the expression shown in Eq.(\ref{eq:frequency_shifts}), 
\begin{equation}
\begin{split}
\delta \omega_\mrm{CFS} & =  n_\mrm{bg} v \frac{\pi}{k^2} \sum_\ell (2\ell+1) \left( \sin 2\phi_{\ell,e} - \sin 2\phi_{\ell,g} \right) \\
& + n_\mrm{bg} v \frac{4 \pi}{k^2}\sum_\ell (2\ell+1) |\sin \phi_{\ell,g} \sin \phi_{\ell,e}| \sin(\phi_{\ell,e} - \phi_{\ell,g}).
\end{split}
\end{equation}

\subsection*{C. Excitation of the $J=0 \leftrightarrow J=2$ rotational transition in H$_2$ during collisions}
The transition probability between the $J=0$ and $J=2$ rotational states in the ground vibrational state in H$_2$ was calculated for the radial component of the electric field gradient. To the potentials $V_\alpha(r)$ shown in Fig.\ \ref{fig:h2-pecs}, we added the centrifugal potential $\frac{\ell(\ell + 1)}{2 \mu r^2}$, and numerically calculated the classical trajectory for the ion-molecule separation $r(t)$ as a function of time. The resulting time-dependent radial electric field gradient, $\frac{\partial \Esca_r}{\partial r} = -\frac{e}{2 \pi \epsilon_0 \, r^3}$, was used to obtain the time-dependent perturbation to the Hamiltonian for the rotational states, $H_\mrm{efg} = Q \,  \frac{\partial \Esca_r}{\partial_r}$. Here $Q = 0.97 \, ea_0^2$ is the electric quadrupole moment matrix element between the $J=0$ and $J=2$ states \cite{Poll1978}. 

\begin{figure}
        \includegraphics[width=0.6\columnwidth]{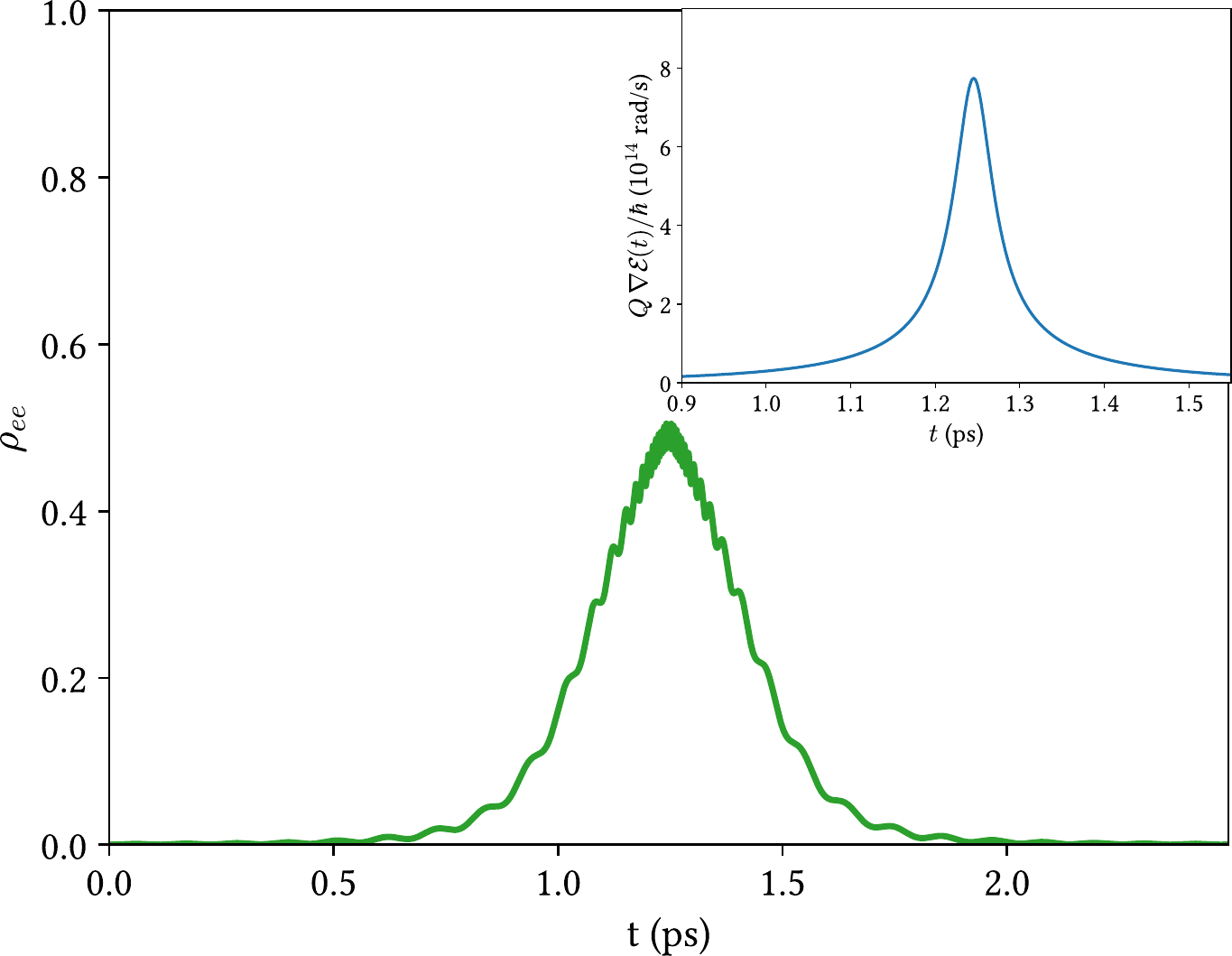}
    \caption{Calculated transition probability for a hydrogen molecule from $\ket{J=0} \to \ket{J=2,m_J=0}$ in the $^1S_0$ PEC, for a classical trajectory with $\ell=0$ and 300 K of collision energy. The inset shows the calculated variation of the transition Rabi frequency, using the electric field gradient experienced by the hydrogen molecule as it collides with the ion. The rotational state population transferred due to the collision is the value of $\rho_{ee}$ at large times.}
    \label{fig:quadrupole_transition}
\end{figure}

Treating the $\ket{J=0}$ and $\ket{J=2,m_J=0}$ rotational states (where $m_J$ is quantized along the collision axis) within the ground vibrational state as a two-level system with energy separation $E_\mrm{rot} = h \times 8.9$ THz, we numerically solved the Schr\"{o}dinger equation for this system with the quadrupole interaction Hamiltonian $H_\mrm{efg}(t)$. The resulting probability for population transfer between $J=0 \leftrightarrow J=2$ was studied for a range of collision energies (between 4-400 K) and partial waves (see Fig.\ \ref{fig:quadrupole_transition}). The largest probability was obtained for high-energy and low-partial-wave collisions as expected, and never exceeded $\sim$ 1\%. The transition probabilities are low because the collision occurs slowly compared to the timescale for rotations of the molecule: the energy levels of the molecule are adiabatically shifted by the electric field gradient from the ion, and there is essentially no population transfer.

\end{document}